\documentclass[aps,prb,reprint,amsmath,amssymb,showpacs]{revtex4-1}

\usepackage{graphicx}
\usepackage{dcolumn}
\usepackage{bm}
\begin{document}


\title{Mixing of odd- and even-frequency pairings \\in strongly correlated electron systems under magnetic field}


\author{Kiyu Fukui}
\affiliation{Department of Physics, The University of Tokyo, Bunkyo, Tokyo 113-0033, Japan}

\author{Yusuke Kato}
\affiliation{Department of Basic Science, The University of Tokyo, Meguro, Tokyo 153-8902, Japan}

\date{\today}

\begin{abstract}
Even- and odd-frequency superconductivity coexist due to broken time-reversal symmetry under magnetic field. In order to describe this mixing, we extend the linearized Eliashberg equation for the spin and charge fluctuation mechanism in strongly correlated electron systems. We apply this extended Eliashberg equation to the odd-frequency superconductivity on a quasi-one-dimensional isosceles triangular lattice under in-plane magnetic field and examine the effect of  the even-frequency component.
\end{abstract}


\maketitle

\indent Odd-frequency superconductivity was originally suggested by Berezinskii in the context of the $^3$He superfluidity in 1974~\cite{Berezinskii1974}. A lot of models showing bulk odd-frequency superconductivity have been studied since the 1990s~\cite{Kirkpatrick1991, Balatsky1992, Emery1992, Belitz1992, Abrahams1993, Coleman1993, Balatsky1993, Vojta1999, Coleman1995, Vojta1999, Fuseya2003a, Shigeta2009, Hotta2009,  Shigeta2011, Kusunose2011, Yanagi2012, Shigeta2012, Harada2012, Shigeta2013, Hoshino2014, Hoshino2014a, Funaki2014, Otsuki2015, Aperis2015}. In addition,  induced odd-frequency Cooper pairs near  surface or interface also have been much investigated~\cite{Bergeret2005, Tanaka2007, Higashitani2012, Tanaka2012, Matsumoto2013}. The odd-frequency superconductivity in bulk materials had been considered to be thermodynamically unstable~\cite{Heid1995} for many years. However, it was proved that bulk odd-frequency superconductivity can be stable by the path-integral formalism~\cite{Belitz1999, Solenov2009, Kusunose2011a}.\\ \indent If there are inversion symmetry and time-reversal symmetry, there are four types of superconductivity: even-frequency spin-singlet even-parity, even-frequency spin-triplet odd-parity, odd-frequency spin-singlet odd-parity, and odd-frequency spin-triplet even-parity. However, it is well known that parity-mixed superconductivity emerges under broken inversion symmetry~\cite{Gorkov2001}. In addition, if we take account of the frequency dependence of the gap functions, even-frequency and odd-frequency states also mix because of broken inversion symmetry~\cite{Yada2009, Shigeta2013b}. Under broken time-reversal symmetry, a similar phenomenon occurs.  The parity mixing does not occur but even- and odd-frequency superconductivity mix with inversion symmetry and broken time-reversal symmetry. In earlier studies~\cite{Matsumoto2012, Kusunose2012, Matsumoto2013a, Kashiwagi2015, Kashiwagi2016}, phonon-mediated even-frequency spin-singlet $s$-wave superconductivity is examined, where odd-frequency spin-triplet ($S_z=0$) $s$-wave minor component mixes with this superconductivity under magnetic field.\\
\indent In the present study, we consider the low-dimensional strongly correlated electron systems under in-plane magnetic fields, in order to examine this mixing effect on odd-frequency anisotropic superconductivity. In those systems, we can avoid the pair-breaking due to orbital magnetism and they are more relevant to real materials~\cite{Uji2001, Shinagawa2007}; in reality, possibility of odd-frequency pairing was discussed~\cite{Pratt2013} in one of organic conductors ($\mathrm{(TMTSF)_2ClO_4}$).

\begin{figure}[b]
\includegraphics[width=6cm]{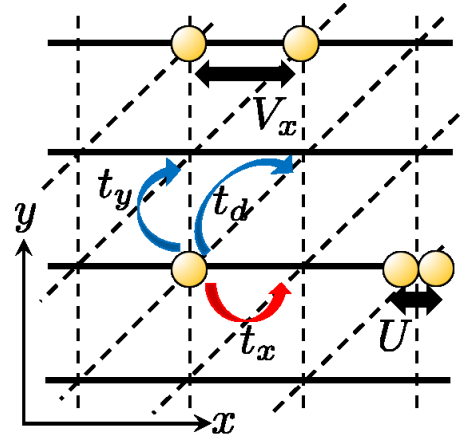}
\caption{\label{fig:lattice} (Color online) The lattice model considered in this study.}
\end{figure}

\indent We start with a single-band extended Hubbard model on a triangular lattice as shown in Fig.~\ref{fig:lattice}, where $t_x$, $t_y$, and $t_d$ are hopping integrals in the $x$, $y$, and diagonal directions, respectively~\cite{Shigeta2009, Shigeta2011}. We consider the effect of magnetic field on the superconductivity through the Zeeman term. We assume in-plane magnetic field and neglect the orbital effect of the electrons. In the present study, we focus on spatially uniform states. We set $k_{\mathrm{B}}=\hbar=1$ and the lattice constant is unity.
The Hamiltonian is given by
\begin{equation}
\begin{split}
\mathcal{H}=&-\sum_{\langle i, j\rangle, \sigma}(t_{ij}c^\dagger_{i\sigma}c_{j\sigma}+\mathrm{H.c.})-h\sum_{i,\sigma}\mathrm{sgn}(\sigma)c^\dagger_{i\sigma}c_{i\sigma}\\
&+U\sum_i n_{i\uparrow}n_{i\downarrow}+\sum_{\langle i,j\rangle}V_{ij}n_i n_j,
\end{split}
\end{equation}
where $t_{ij}$ is the hopping integral between sites $i$ and $j$, and $\langle i, j\rangle$ is a pair of the nearest neighbors. $c_{i\sigma}$ ($c^\dagger_{i\sigma}$) is an annihilation (creation) operator for the electron with spin $\sigma$ on site $i$. $n_{i\sigma}=c^\dagger_{i\sigma}c_{i\sigma}$, $n_i=n_{i\uparrow}+n_{i\downarrow}$ and $\mathrm{sgn}(\sigma)=+1$ ($-1$) for $\sigma=\uparrow$ ($\downarrow$). $h$ is the magnitude of the magnetic field. Here we consider that spins and momenta of the electrons have the same direction for simplicity. $U$ and $V_{ij}$ stand for the on-site and off-site Coulomb repulsions, respectively and $V_{ij}$ acts pairs on the nearest neighbor of electrons in the $x$ direction ($V_x$). Here we set $t_y=t_d=0.1t_x$, $U=1.6t_x$ and half-filling. The dispersion relation of the electron with spin $\sigma$ is given by
$\varepsilon_{\bm{k}\sigma}=-2t_x\cos k_x-2t_y\cos k_y-2t_d\cos(k_x+k_y)-\mathrm{sgn}(\sigma)h$ and the momentum dependence of the off-site Coulomb repulsion is given by $V(\bm{q})=2V_x\cos q_x$ where $\bm{k}=(k_x, k_y)$ and $\bm{q}=(q_x, q_y)$. The nesting vector is $(\pi, \pi/2)$. 
This model shows odd-frequency spin-singlet $p$-wave (OS$p$) superconductivity and odd-frequency spin-triplet $s$-wave (OT$s$) superconductivity in zero magnetic field as shown in earlier studies~\cite{Shigeta2009, Shigeta2011}. Owing to the disconnected Fermi surface in this model, OS$p$ as well as OT$s$ can be realized without the gap nodes. Further, the Coulomb interaction becomes effectively attractive for \textit{s-wave triplet} pairing. These are the reasons why this model has odd-frequency superconductivity with those pairings. We use this model to examine the effect of in-plane magnetic field on the superconductivity in the low-dimensional strongly correlated electron systems. This effect has not been addressed so far in the references~\cite{Shigeta2009, Shigeta2011} but important as we remarked in the introduction.
\\
\indent We extend the linearized Eliashberg equation to the cases where even- and odd-frequency pairs (i.e. singlet and part of triplet pair with $z$-component of spin $S_z$ being zero) coexist in magnetic field. Strictly speaking, our system of equations should be called ^^ ^^ linearized gap equations” because the present scheme is not conservation approximation. In the present study, however, we use the term the linearized Eliashberg equation in the sense that we take account of the strong coupling effect in the gap equation, following earlier references~\cite{Shigeta2011, Yanagi2012, Shigeta2013}.\\
\indent The $S_z=\pm 1$ pairs do not mix with singlet pairs and the linearized Eliashberg equations for them are given by
\begin{equation} 
\lambda\Delta_{\sigma\sigma}(k)=-\frac{T}{N}\sum_{k'}\Gamma^{\sigma\sigma}_{k-k'}G^0_{\sigma\sigma}(k')G^0_{\sigma\sigma}(-k')\Delta_{\sigma\sigma}(k'),
\label{eq:eliashberg_parallel}
\end{equation}
where $\sigma=\uparrow$ ($\downarrow$) for $S_z=+1$ ($S_z=-1$) pairings, $N$ is the number of sites ($=$ the number of unit cells), $T$ is the temperature. $k\equiv(\mathrm{i}\omega_n, \bm{k})$, where $\omega_n = (2n+1)\pi T$ with an integer $n$ is a fermionic Matsubara frequency. $G^0_{\sigma\sigma}(k)=(\mathrm{i}\omega_n-\varepsilon_{\bm{k}\sigma}+\mu)^{-1}$ is the Green's function with  the chemical potential $\mu$. $\lambda$ is the eigenvalue for  the gap function $\Delta_{\sigma\sigma}(k)$. The effective pairing interactions mediated by spin and charge fluctuations $\Gamma^{\sigma\sigma}_q$ within the random-phase approximation (RPA) are~\cite{Aizawa2008, Aizawa2009, Aizawa2009a}
\begin{equation}
\begin{split}
\Gamma^{\sigma\sigma}_{q}&=V(\bm{q})-[U+V(\bm{q})]^2\chi^{\bar{\sigma}\bar{\sigma}}(q)-V(\bm{q})^2\chi^{\sigma\sigma}\\
&-2[U+V(\bm{q})]V(\bm{q})\chi^{\sigma\bar{\sigma}}(q),
\end{split}
\end{equation}
where $\bar{\sigma}$ is the spin in the opposite direction to $\sigma$ and $q=(\mathrm{i}\nu_m,\ \bm{q})$, where $\nu_m=2m\pi T$ with an integer $m$ is a bosonic Matsubara frequency. The susceptibilities $\chi^{\sigma\sigma}$ and $\chi^{\sigma\bar{\sigma}}$ are 
\begin{equation}
\chi^{\sigma\sigma}(q)=[1+V(\bm{q})\chi^{\bar{\sigma}\bar{\sigma}}_0(q)]\chi^{\sigma\sigma}_0(q)/D(q),
\end{equation}
and 
\begin{equation}
\chi^{\sigma\bar{\sigma}}(q)=-[U+V(\bm{q})]\chi^{\sigma\sigma}_0(q)\chi^{\bar{\sigma}\bar{\sigma}}_0(q)/D(q).
\end{equation}
Their denominator $D(q)$ is given by 
\begin{equation}
\begin{split}
D(q)&=[1+V(\bm{q})\chi^{\sigma\sigma}_0(q)][1+V(\bm{q})\chi^{\bar{\sigma}\bar{\sigma}}_0(q)]\\
&-[U+V(\bm{q})]^2\chi^{\sigma\sigma}_0(q)\chi^{\bar{\sigma}\bar{\sigma}}_0(q),
\end{split}
\end{equation}
and the longitudinal irreducible susceptibilities are
\begin{equation}
\chi^{\sigma\sigma}_0(q)=\frac{1}{N}\sum_{\bm{k}}\frac{f(\varepsilon_{\bm{k}+\bm{q}\sigma})-f(\varepsilon_{\bm{k}\sigma})}{\mathrm{i}\nu_m-\varepsilon_{\bm{k}+\bm{q}\sigma}+\varepsilon_{\bm{k}\sigma}},
\end{equation}
with the distribution function $f(\varepsilon_{\bm{k}\sigma})=[\mathrm{e}^{(\varepsilon_{\bm{k}\sigma}-\mu)/T}+1]^{-1}$.
The linearized Eliashberg equation for $S_z=0$ pairings is given by
\begin{widetext}
\begin{equation}
\lambda
\begin{pmatrix}
\Delta_{\mathrm{S}}(k)\\ \\\Delta_{\mathrm{T^0}}(k)\end{pmatrix}
=-\frac{T}{N}\sum_{k'}
\begin{pmatrix}
\Gamma^{\mathrm{S}}_{k-k'}f^+_{k'}-\Gamma^-_{k-k'}f^-_{k'} & &
\Gamma^{\mathrm{S}}_{k-k'}f^-_{k'}-\Gamma^-_{k-k'}f^+_{k'}\\ \\
\Gamma^{\mathrm{T^0}}_{k-k'}f^-_{k'}+\Gamma^-_{k-k'}f^+_{k'} & &
\Gamma^{\mathrm{T^0}}_{k-k'}f^+_{k'}+\Gamma^-_{k-k'}f^-_{k'}
\label{eq:eliashberg_antiparallel}
\end{pmatrix}
\begin{pmatrix}
\Delta_{\mathrm{S}}(k')\\ \\\Delta_{\mathrm{T^0}}(k')\end{pmatrix},
\end{equation}
\end{widetext}
where $\Gamma^{\mathrm{S}}_q$, $\Gamma^{\mathrm{T^0}}_q$ are the effective pairing interactions for singlet pairs
\begin{equation}
\begin{split}
\Gamma^{\mathrm{S}}_q&=U+V(\bm{q})+\frac{U^2}{2}\chi^{zz}_{\mathrm{s}}(q)-\frac{1}{2}[U+2V(\bm{q})]^2\chi_{\mathrm{c}}(q)\\
&+\frac{U^2}{2}[\chi^{-+}_{\mathrm{s}}(q)+\chi^{+-}_{\mathrm{s}}(q)],
\end{split}
\end{equation}
and for $S_z=0$ triplet pairs
\begin{equation}
\begin{split}
\Gamma^{\mathrm{T^0}}_q&=V(\bm{q})+\frac{U^2}{2}\chi^{zz}_{\mathrm{s}}(q)-\frac{1}{2}[U+2V(\bm{q})]^2\chi_{\mathrm{c}}(q)\\
&-\frac{U^2}{2}[\chi^{-+}_{\mathrm{s}}(q)+\chi^{+-}_{\mathrm{s}}(q)],
\end{split}
\end{equation}
respectively.
In addition, $\Gamma^-_q$ and $f^{\pm}_k$ are given by
\begin{equation}
\Gamma^-_q=\frac{U^2}{2}[\chi^{-+}_{\mathrm{s}}(q)-\chi^{+-}_{\mathrm{s}}(q)],
\end{equation}
and
\begin{equation}
f^{\pm}_k=\frac{1}{2}[G^0_{\uparrow\uparrow}(k)G^0_{\downarrow\downarrow}(-k)\pm G^0_{\downarrow\downarrow}(k)G^0_{\uparrow\uparrow}(-k)],
\end{equation}
respectively. The longitudinal spin susceptibility and charge susceptibility are 
$\chi^{zz}_{\mathrm{s}}(q)=\frac{1}{2}[\chi^{\uparrow\uparrow}(q)-\chi^{\uparrow\downarrow}(q)-\chi^{\downarrow\uparrow}(q)+\chi^{\downarrow\downarrow}(q)]$,
and 
$\chi_{\mathrm{c}}(q)=\frac{1}{2}[\chi^{\uparrow\uparrow}(q)+\chi^{\uparrow\downarrow}(q)+\chi^{\downarrow\uparrow}(q)+\chi^{\downarrow\downarrow}(q)]$,
respectively. The transverse spin susceptibilities within the RPA are 
\begin{equation}
\chi^{-+\ (+-)}_{\mathrm{s}}(q)=\frac{\chi^{-+\ (+-)}_{\mathrm{s}0}(q)}{1-U\chi^{-+\ (+-)}_{\mathrm{s}0}(q)},
\end{equation}
where irreducible transverse susceptibilities are
\begin{equation}
\chi^{-+\ (+-)}_{\mathrm{s}0}(q)=\frac{1}{N}\sum_{\bm{k}}\frac{f(\varepsilon_{\bm{k}+\bm{q}\uparrow(\downarrow)})-f(\varepsilon_{\bm{k}\downarrow(\uparrow)})}{\mathrm{i}\nu_m-\varepsilon_{\bm{k}+\bm{q}\uparrow(\downarrow)}+\varepsilon_{\bm{k}\downarrow(\uparrow)}}. 
\end{equation}
We ignore the effects of the off-site Coulomb repulsion in the transverse susceptibilities; it is difficult to treat them in RPA~\cite{Aizawa2008, Aizawa2009, Aizawa2009a} because of the ladder-type diagrams. 
In Eq.~(\ref{eq:eliashberg_antiparallel})  the singlet gap function $\Delta_{\mathrm{S}}$ and the $S_z=0$ triplet gap function $\Delta_{\mathrm{T}^0}$ are mixed. In Eq.~(\ref{eq:eliashberg_antiparallel}), the functions whose superscripts are minus signs are zero if $h=0$, and the matrix in the equation becomes diagonal. Therefore the singlet component and the triplet component are decoupled and reduced to the conventional linearized Eliashberg equations in RPA.\\
\indent The linearized Eliashberg equations shown above are eigenequations. We solve these equations numerically and obtain eigenvalues $\lambda$ and normalized gap functions [$\sum_k\lvert \Delta(k)\rvert^2=1$] . Superconducting instability occurs, when the largest real eigenvalue reaches unity. In this paper, we take $N=64\times64$ $\bm{k}$-meshes and 1025 Matsubara frequencies from $-(2N_{\mathrm{f}}+1)\pi T$ to $(2N_{\mathrm{f}}+1)\pi T$ with $N_{\mathrm{f}}=512$. We also calculate eigenvalues for some sets of parameters in the cases of $N=32\times32$, $N_{\mathrm{f}}=256$ and $N=128\times64$, $N_{\mathrm{f}}=1024$ and check that eigenvalues do not change drastically.\\
\begin{figure}
\includegraphics[width=8cm]{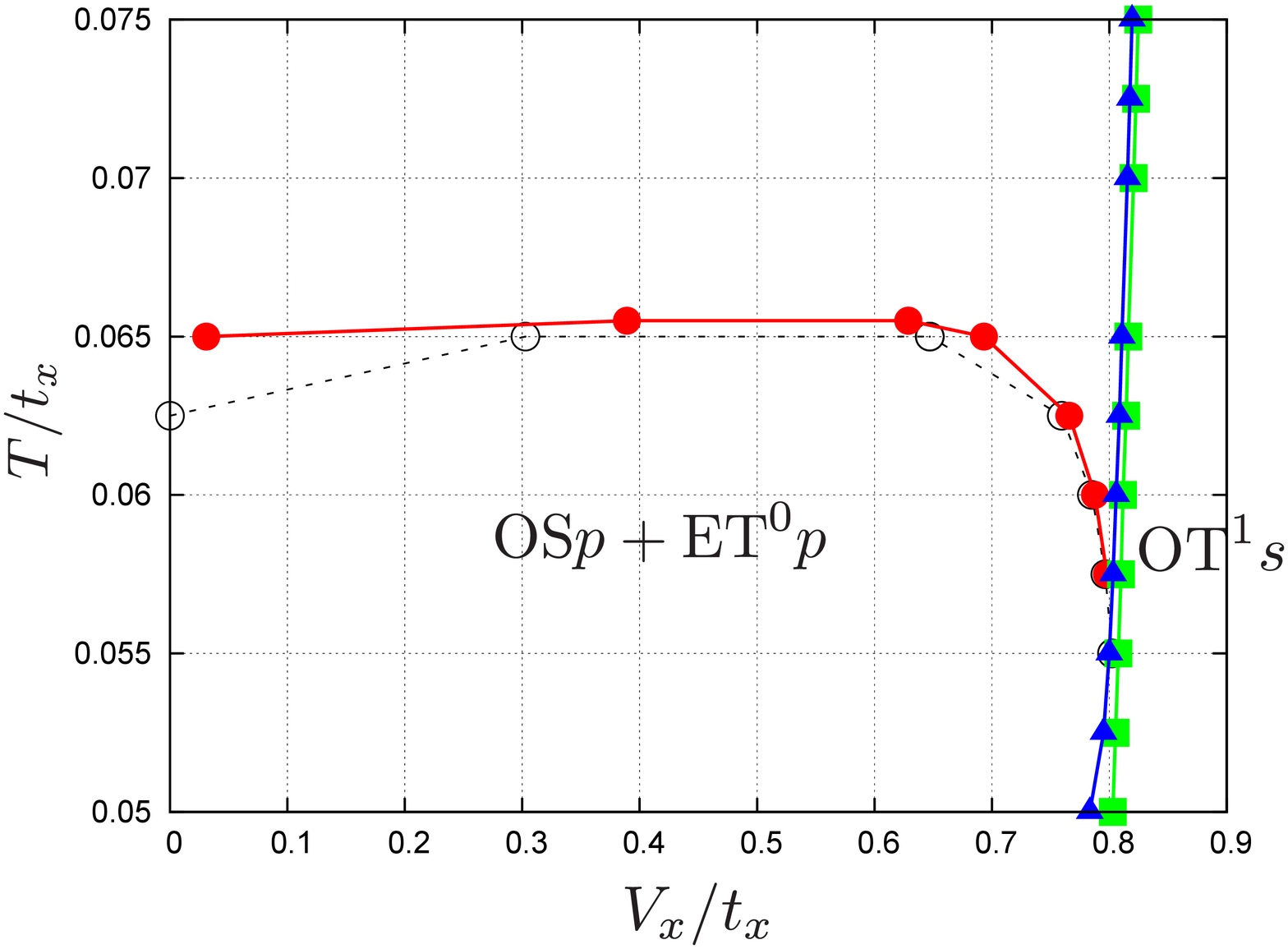}
\caption{\label{fig:T_V_dig} (Color online) Phase boundary on the $T\mathchar`-V_x$ plane. We set $h/t_x=0.025$. The filled red circles and blue triangles show the transition temperatures of OS$p$+ET$^0p$ and OT$^1s$, respectively. The green squares show the temperatures where the longitudinal spin susceptibility and the charge susceptibility diverge in normal state. The black open circles show the transition temperature of OS$p$ calculated by neglecting ET$^0p$ component. Lines are eye guides.}
\end{figure}
\begin{figure}
\includegraphics[width=8cm]{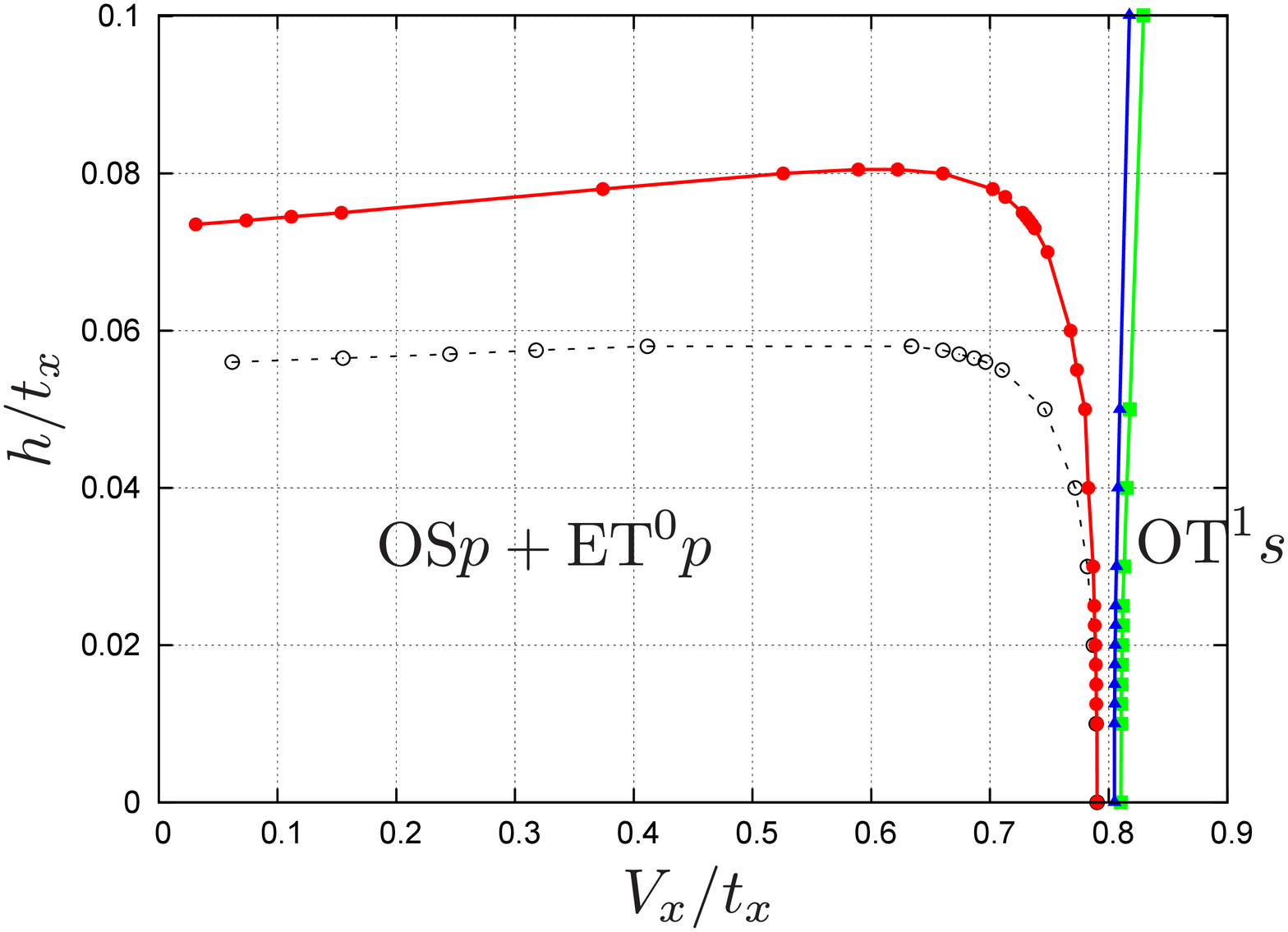}
\caption{\label{fig:h_V_dig} (Color online) Phase boundary on the $h\mathchar`-V_x$ plane. We set $T/t_x=0.06$. The red circles and blue triangles show the critical fields of OS$p$+ET$^0p$ and OT$^1s$, respectively. The green squares show the magnetic fields where the longitudinal spin susceptibility and the charge susceptibility diverge in normal state. The black open circles show the critical fields of OS$p$ calculated by neglecting ET$^0p$ component. Lines are eye guides.}
\end{figure}
\begin{figure}
\includegraphics[width=8cm]{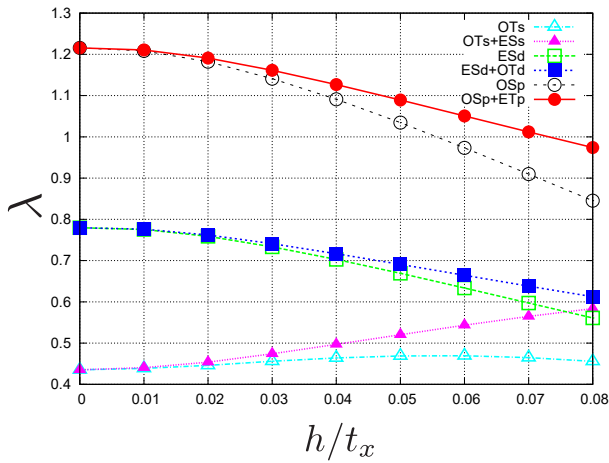}
\caption{\label{fig:eigenvalues} (Color online) $h$ dependence of the eigenvalues. We set $V_x/t_x=0$ and $T/t_x=0.06$ here.}
\end{figure}
\indent First, we set $h=2.5 t_x$ and calculate eigenvalues of the linearized Eliashberg equations (\ref{eq:eliashberg_parallel}) and (\ref{eq:eliashberg_antiparallel}) changing $T$ and $V_x$. The results are shown in Fig.~\ref{fig:T_V_dig}. In the case of $h\neq 0$, eigenvalues of Eq. (\ref{eq:eliashberg_antiparallel}) are not necessarily real~\cite{Matsumoto2012, Kusunose2012}, but the largest eigenvalues are always real in this study.
The red filled circles show temperatures where the largest eigenvalue of OS$p$ major component with ET$^0p$ minor component reaches unity. ^^ ^^ ET$^0p$" stands for even-frequency $S_z=0$ spin-triplet $p$-wave. When $h=0$, this superconductivity has pure OS$p$ symmetry. The ET$^0p$ component mixes with this OS$p$ component in magnetic field. Similarly the blue triangles show temperatures where the largest eigenvalues of OT$^1s$ superconductivity reaches unity. ^^ ^^ OT$^1s$" represents odd-frequency $S_z=1$ spin-triplet $s$-wave. These symbols show transition temperatures ($T_{\mathrm{c}}$) of each superconductivity. The momentum and frequency dependence of each gap function is shown in the supplemental materials. The lines connecting symbols are guides for eye. The green squares show temperatures where the longitudinal spin susceptibility and the charge susceptibility diverge in normal state. In RPA calculation in finite magnetic field, the denominator of the longitudinal spin susceptibility and that of the charge susceptibility are the same. Hence they diverge at the same point. We found that there is no parameter region where spin density wave and charge density wave states have higher transition temperatures than any superconducting states. The off-site Coulomb repulsion ($V_x$) enhances charge fluctuation. Charge fluctuation enhances triplet superconductivity while it suppresses singlet superconductivity. On the other side, spin susceptibility enhances both singlet and triplet superconductivity. Therefore singlet superconductivity is suppressed and triplet superconductivity is enhanced near the green line. In order to see how the mixing of the minor component affects the superconducting stability, we calculate $T_{\mathrm{c}}$ by solving eigenvalues of Eq.~(\ref{eq:eliashberg_antiparallel}) with $\Delta_{\mathrm{T}^0}=0$. The result is shown by the black open circles in Fig.~\ref{fig:T_V_dig}. It seems that the minor ET$^0p$ component does not affect or  raises $T_{\mathrm{c}}$ a little. We should note that pure OS$p$ superconductivity alone cannot be the solution of the linearized Eliashberg equation under magnetic field and the mixing of the ET$^0p$ component is inevitable~\cite{Matsumoto2012, Kusunose2012}, as you can see Eq.~(\ref{eq:eliashberg_antiparallel}). \\
\indent Next, we set $T=0.06 t_x$ and calculate eigenvalues. The results are shown in Fig.~\ref{fig:h_V_dig}. As in Fig.~\ref{fig:T_V_dig}, the filled red circles and blue triangles show the magnetic fields where the eigenvalues of OS$p$+ ET$^0p$ superconductivity and OT$^1s$ superconductivity, respectively, reach unity.  The green squares show the magnetic fields where the longitudinal spin susceptibility and the charge susceptibility diverge in normal state. As the above, to examine how the mixing of the minor component affects the superconducting stability, we solve Eq.~(\ref{eq:eliashberg_antiparallel}) with $\Delta_{\mathrm{T}^0}=0$. The result is shown by the black open circles.  The dependence of all lines on $V_x$ is roughly the same as Fig.~\ref{fig:T_V_dig}. We can see that mixing of the minor component makes superconducting phase more stable.\\
\indent In the preceding study~\cite{Matsumoto2012, Matsumoto2013a}, Einstein phonon-mediated ES$s$ superconductivity with OT$^0s$ minor component in magnetic field is studied and the minor OT$^0s$ component suppresses $T_{\mathrm{c}}$. Here ^^ ^^ ES$s$" and ^^ ^^ OT$^0s$" mean even-frequency spin-singlet $s$-wave and  odd-frequency $S_z=0$ spin-triplet $s$-wave, respectively. There are main two differences between the previous and the present studies. The first difference is that in the Einstein phonon case (earlier study) the major component is even-frequency, on the other hand  in the spin and charge fluctuation case (present study)  the major component is odd-frequency. The second difference is the signs of the effective pairing interactions. The pairing interaction from the electron-phonon interaction is attractive. However the pairing interaction for OS$p$ in the present study is repulsive. To examine the effects due to these two differences, we focus on the second and third largest eigenvalues of Eq.~(\ref{eq:eliashberg_antiparallel}). The solution of Eq.~(\ref{eq:eliashberg_antiparallel}) with the second largest eigenvalue is  ES$d$ superconductivity with OT$^0d$ minor component. ^^ ^^ ES$d$" and ^^ ^^ OT$^0d$" represent even-frequency spin-singlet $d$-wave and odd-frequency $S_z=0$ spin-triplet $d$-wave, respectively. Furthermore, the solution with the third largest eigenvalue is OT$^0s$ superconductivity with ES$s$ minor component.  We show magnetic field dependence of these eigenvalues and eigenvalues calculated by ignoring the minor components in Fig.~\ref{fig:eigenvalues}. In all cases the minor components enhance superconductivity. Even though the major component is even-frequency, the minor odd-frequency component does not suppress superconductivity in this study. The effective pairing interactions for triplet pairing are attractive in the spin and charge fluctuation mechanism. In addition the minor component also enhance superconductivity, though the effective pairing interaction for OT$^0s$ superconductivity is attractive. From these results above, the two differences between earlier study and present study are not essential. We apply Eq.~(\ref{eq:eliashberg_antiparallel}) to ES$d$ superconductivity on a model for cuprate high-temperature superconductors~\cite{Yanase2003} for confirmation. The OT$^0d$ minor component does not suppress ES$d$ superconductivity in cuprates model in $h\neq 0$. The difference between the electron-phonon case and the spin and charge fluctuation case may be caused by something different from what are mentioned above. \\
\indent In summary, we have extended the linearized Eliashberg equations for the spin and charge fluctuation-mediated superconductivity to examine even- and odd-frequency mixing under magnetic field, applied them to a model of a quasi-one-dimensional isosceles trianglar lattice, and solved these equations. As the results, the minor components enhance superconductivity unlike the results of the earlier study for phonon-mediated superconductivity. A future work is to extend the present study on the basis of realistic electronic structure. Possibility of the spatially inhomogeneous state should be also considered. Further comparison with the electron-phonon systems in the difference of mixture effect of even- and odd-frequency pairing on the superconducting properties remains as a theoretical issue.\\
\\
\indent We would like to thank H. Kusunose, Y. Yanagi, S. Hoshino, S. Kurihara, D. Yamamoto, K. Masuda, S. Goto, and Y. Masaki for helpful discussions and comments. This work was supported by JSPS KAKENHI Grant Number 15K05160. K. F. was also supported by the JSPS through Program for Leading Graduate Schools (MERIT).
\bibliographystyle{apsrev4-1}
\bibliography{library}

\end{document}